# Sub-60 mV/decade switching via high energy electrons tunneling in nanoscale gallium nitride field-effect transistors

Peng Cui, Guangyang Lin, Jie Zhang, and Yuping Zeng[a)]

Department of Electrical and Computer Engineering, University of Delaware, Newark, DE 19716, USA

Novel devices such as tunneling field-effect transistors (FETs) and ferroelectric FETs have been demonstrated to break the subthreshold swing (SS) limit with sub-60 mV/decade switching for further low voltage/low power applications. In this paper, SS of sub-60 mV/dec was firstly observed in InAlN/GaN high electron mobility transistors (HMETs): an average SS of 30 mV/dec over three orders of magnitude in drain-source ($I_{ds}$) and a minimum point-by-point SS of 15 mV/dec were achieved in the InAlN/GaN HEMTs with gate length ($L_g$) of 40 nm. It is found that SS decreases with drain-source voltage ($V_{ds}$) as well as $L_g$, and falls below 60 mV/dec when $L_g < 100$ nm. The decrease of SS as the device dimension scales down is attributed to the tunneling of high energy electrons from channel to the surface. The SS decreasing with the nanoscale gate length shows the great potential of the InAlN/GaN HEMTs to be applied in future logic switches.

---

[a)] Author to whom correspondence should be addressed. Electronic mail: yzeng@udel.edu.

Boltzmann distribution dictates that, to affect an order of magnitude increase in the drain current, the gate voltage needs to change by at least 60 mV at 300 K.[1-3] This means that the subthreshold swing (SS) in conventional field effect transistors (FETs) cannot fall below 60 mV/dec at room temperature, which will hinder the devices for further low voltage/low power applications. Novel devices such as tunneling FETs[4,5] and ferroelectric FETs[6,7] have been demonstrated to break this SS limit with sub-60 mV/decade switching.

Due to the high electron velocity and large bandgap, gallium nitride high electron mobility transistors (HEMTs) have been demonstrated for the high-frequency and high-power applications in the last two decades.[8-10] InAlN/GaN metal-insulator-semiconductor HEMTs (MIS-HEMTs) with SS below 60 mV/dec has been reported by several groups.[11-14] There were several similar characteristics reported in these devices: 1) gate dielectrics are deposited; 2) gate length ($L_g$) is 1~3 μm; 3) SS of sub-60 mV/dec was observed with the forward sweep in gate-source voltage ($V_{gs}$, from low to high).

In this work, we report for the first time that sub-60 mV/dec SS can be achieved in GaN HEMTs without gate dielectric deposition. Different from other InAlN/GaN MISHEMTs, the SS of sub-60 mV/dec was observed with a reverse sweep in $V_{gs}$ (from high to low). Additionally, it is found that SS decreases with $L_g$ as well as drain-source voltage ($V_{ds}$), and falls below 60 mV/dec when $L_g < 100$ nm. As the device dimension scales down, the high energy electrons tunnel from channel to the surface dominates, contributing to the decreasing of SS.

**Fig. 1(a)** shows the fabricated InAlN/GaN HEMT. The lattice-matched $In_{0.17}Al_{0.83}$-N/GaN heterostructure was grown by metalorganic chemical vapor deposition (MOCVD) on a Si substrate. The epitaxial layer, from the bottom to the top, consists of a 2-μm undoped GaN buffer layer, a 4-nm $In_{0.12}Ga_{0.88}N$ back barrier layer, a 15-nm GaN channel layer, a 1-nm AlN interlayer, an 8-nm lattice-matched $In_{0.17}Al_{0.83}N$ barrier layer, and 2-nm GaN cap layer. The two-dimensional electron gas (2DEG) electron density and electron mobility determined using Hall measurement were $2.28 \times 10^{13}$ $cm^{-2}$ and 1205 $cm^2/(V \cdot s)$, respectively. Device fabrication process started with mesa isolation by $Cl_2$-based inductively coupled plasma (ICP) etching. Alloyed ohmic contact of Ti/Al/Ni/Au metal stack was then deposited and annealed at 850˚C for 30s in forming gas. Subsequently, sample surface was treated with oxygen plasma treatment to form an oxide layer on the surface[15,16]. Finally, gate electrodes with gate width ($W_g$) of 50 × 2 μm were defined using electron beam lithography and Ni/Au metal stack deposition. The device gate lengths are 40, 70, 100, 150, and 300 nm with source-drain distance ($L_{sd}$) of 2 μm. The device gate lengths are 1, 2, and 3 μm with $L_{sd}$ of 6 μm. The DC current-voltage ($I$–$V$) measurements were carried out by using an Agilent B1500A semiconductor parameter analyzer.

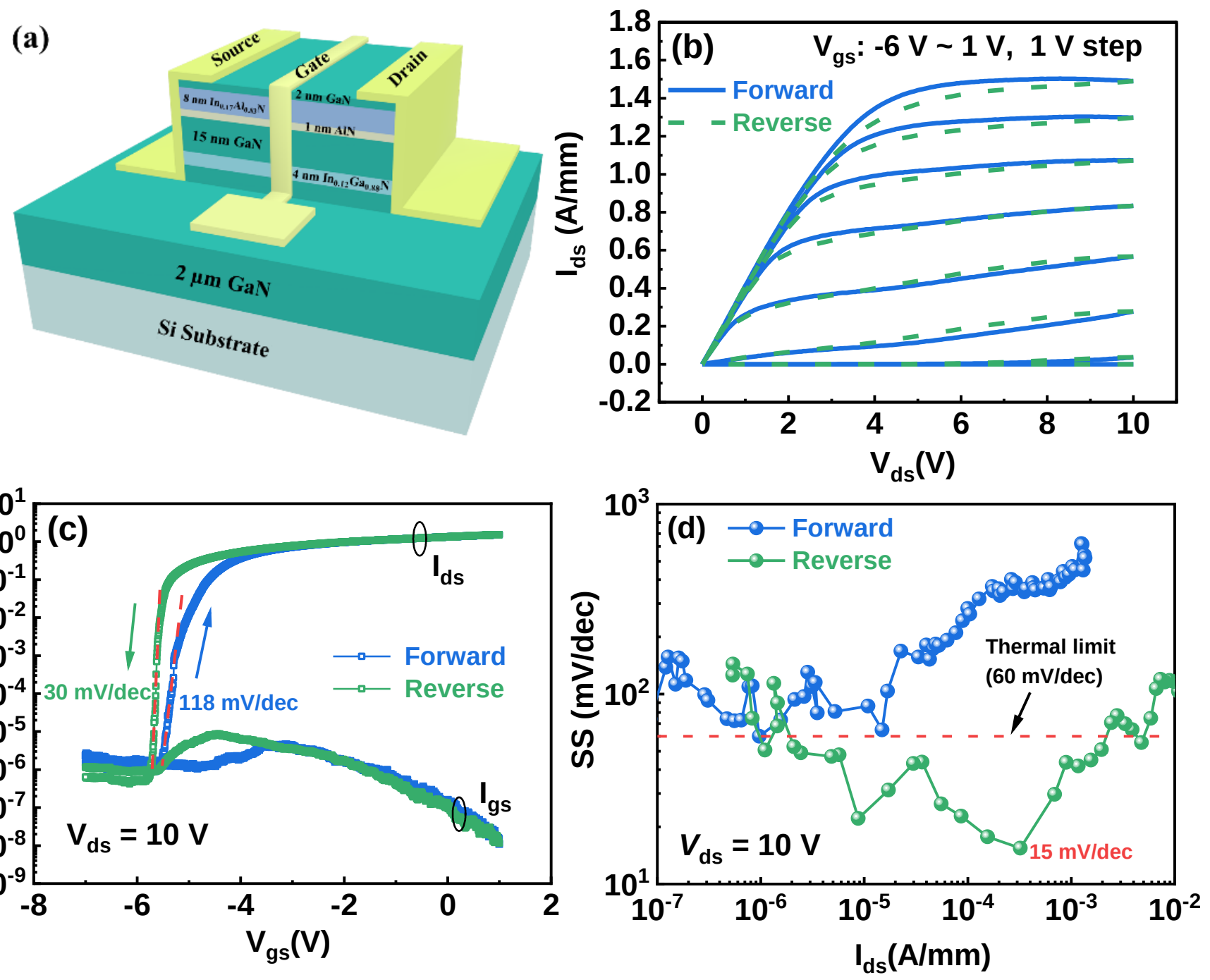

**Fig. 1 (a)** Schematic of the fabricated InAlN/GaN HEMT; **(b)** *I-V* output characteristics of 40-nm InAlN/GaN HEMT with forward/reverse sweep in $V_{ds}$; **(c)** Transfer characteristics and gate leakage current curves of 40-nm InAlN/GaN HEMT at $V_{ds}$ = 10 V with forward/reverse sweep in $V_{gs}$; **(d)** Point-by-point subthreshold swing (SS) vs drain current $I_{ds}$ at $V_{ds}$ = 10 V with forward/reverse sweep in $V_{gs}$.

**Fig. 1(b)** shows the measured *I-V* characteristics of 40-nm InAlN/GaN HEMT with forward/reverse sweep in drain-source voltage ($V_{ds}$) under the same gate-source voltage ($V_{gs}$). Here $V_{gs}$ is swept from -6 V to 1 V and a significant hysteresis of drain-source current ($I_{ds}$) is observed. The on-resistance ($R_{on}$) extracted at $V_{gs}$ = 0 V and $V_{ds}$ between 0 V to 0.5 V are 2.59/2.51 Ω•mm with forward/reverse sweep in $V_{ds}$. **Fig. 1(c)** shows the transfer curves in semi-log scale and gate-source leakage current ($I_{gs}$) of 40-nm InAlN/GaN HEMT at $V_{ds}$ = 10 V with forward/reverse sweep in $V_{gs}$. The on current ($I_{on}$), off current ($I_{off}$), on/off current ($I_{on}/I_{off}$) ratio, and $I_{gs}$ (at $V_{gs}$ = -7V) with forward/reverse sweep in $V_{gs}$ are 1.52/1.52 A/mm, $8.95\times10^{-7}/4.17\times10^{-7}$ A/mm, $1.69\times10^{6}/3.64\times10^{6}$, and 1.49 $\times10^{-6}$/1.16 $\times10^{-6}$ A/mm, respectively. $I_{off}$ decreased and $I_{on}/I_{off}$ increased with reverse sweep in $V_{gs}$. To determine SS, the sweep step of $V_{gs}$ is set at 5 mV. The extracted average SS values over three/four orders of magnitude in $I_{ds}$ are 118/30 mV/dec with forward/reverse sweep in $V_{gs}$. The SS in reverse sweep of 30 mV/dec is far below the thermal limit of 60 mV/dec with a sharp increase of $I_{ds}$. **Fig. 1(d)** presents the point-by-point SS vs $I_{ds}$ extracted from **Fig. 1(c)**. In the subthreshold region, the SS with forward sweep in $V_{gs}$ are all above 60 mV/dec, while SS in reverse sweep of $V_{gs}$ falls below the thermal limit for over 3 decades of $I_{ds}$, with a minimum of 15 mV/dec.

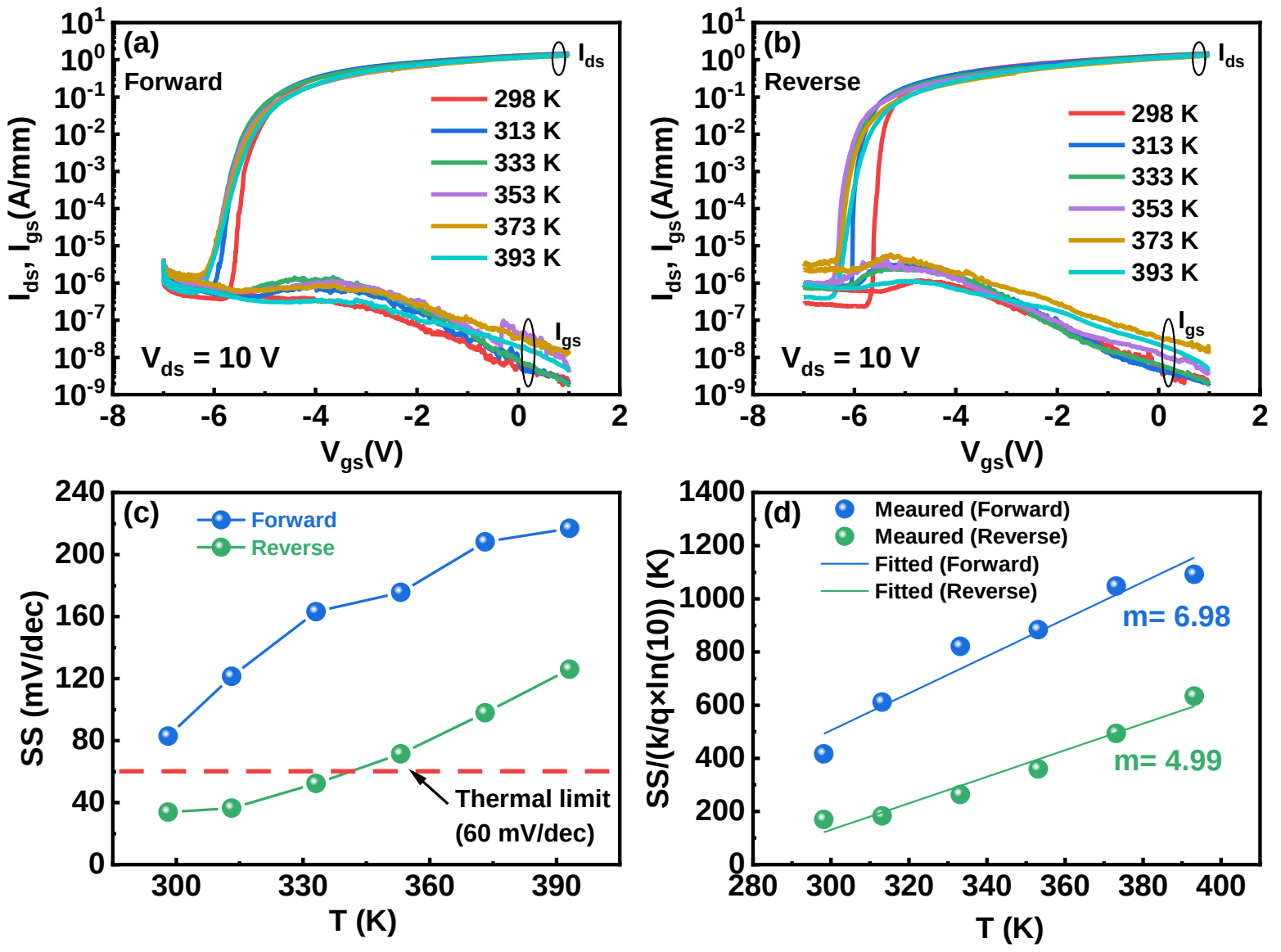

**Fig. 2** Transfer curves in semi-log and gate leakage current at $V_{ds}$ = 10 V of 40-nm InAlN/GaN HEMT under different temperature **(a)** with forward sweep in $V_{gs}$ and **(b)** with reverse sweep in $V_{gs}$. **(c)** The SS as a function of temperature $T$ with forward/reverse sweep in $V_{gs}$. **(d)** The measured and fitted SS/($k$/q ×ln(10)) as a function of temperature T with forward/reverse sweep in $V_{gs}$.

SS can be expressed as[2,3,11]

$$SS = \frac{\partial V_{gs}}{\partial(\log_{10} I_{ds})} = \frac{\partial V_{gs}}{\partial \psi_s} \cdot \frac{\partial \psi_s}{\partial(\log_{10} I_{ds})} = mkT / q * \ln(10), \qquad (1)$$

where $\psi_s$ is the channel potential, $k$ is the Boltzmann's constant, $T$ is the temperature, $q$ is the electron charge. $m = \partial V_{gs}/\partial \psi_s = 1 + C_s/C_{ins}$ is the body factor.[11] $C_s$ and $C_{ins}$ are the GaN channel capacitance and the insulator capacitance, respectively. Here $m$ is usually exceeding one, thus SS features a lower limit of 60 mV/dec when $m$ is 1.[2,3,11] When $m$ < 1, device presents a negative capacitance.[2,3] To obtain the value of $m$, the transfer curves of 40-nm InAlN/GaN HEMT under different temperatures at $V_{ds}$ = 10 V were measured in **Fig. 2(a)** and **(b)**. With the increased temperature, the gate leakage current is almost unchanged, an indication that the tunnel current dominates the gate leakage current. However, the slope of $I_{ds}$-$V_{gs}$ at the subthreshold region decreased. **Fig. 2(c)** shows the extracted average SS under different temperature with the forward/reverse sweep of $V_{gs}$. SS decreased with the increased temperature in both forward/reverse sweep of $V_{gs}$ and started to be higher than 60 mV/dec when T > 333 K with reverse sweep of $V_{gs}$. **Fig. 2(d)** shows the measured and fitted $SS/(k/q \times \ln(10))$ extracted from **Fig. 2(c)**. $m$ values of 4.99/6.98 were obtained with the fitted curves for the forward/reverse sweep of $V_{gs}$. This means the steep SS characteristic with reverse sweep is not due to the negative capacitance effect.

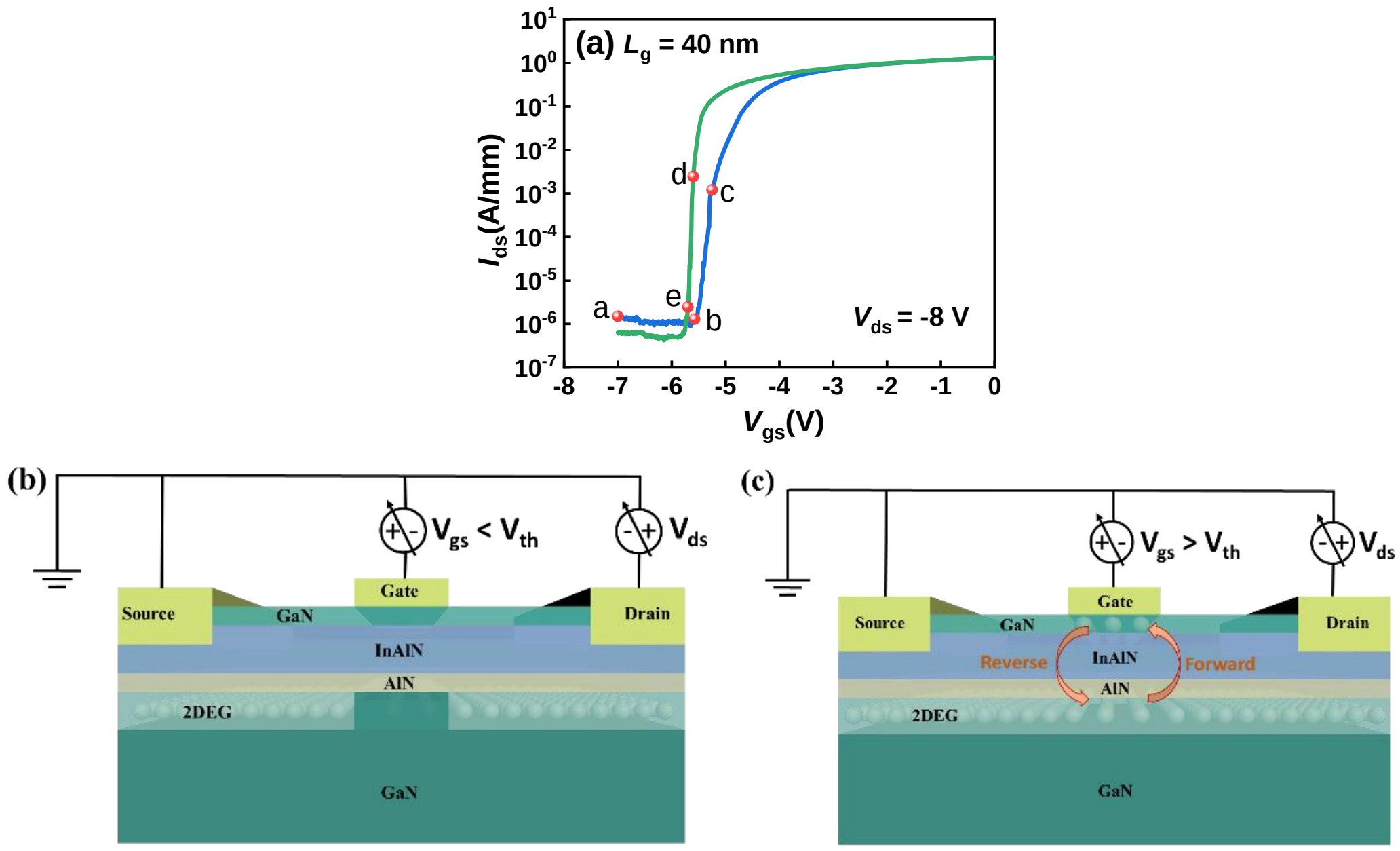

**Fig. 3 (a)** The transfer curves in semi-log at $V_{ds}$ = 10 V; **(b)** Schematic of the gate depletion region with gate voltage $V_{gs}$ < $V_{th}$ (threshold voltage); **(c)** Schematic of the electron injection and release with forward/reverse sweep in $V_{gs}$ when $V_{gs}$ > $V_{th}$.

Another possible reason should come from the interface states between the gate metal and GaN cap layer. These interface states can cause the electron injection and release of 2DEG electrons and the subthreshold characteristics variation with forward/reverse sweep. Due to the short gate length and source/drain distance ($L_g$ of 40 nm and $L_{sd}$ of 1 μm), the electric field between the source-drain channel, especially under the gate region is very high. The 2DEG electrons under the gate region features a high velocity under the high electric field. These high velocity electrons obtain high energy due to the strong electrical field and possibly tunnel through the InAlN barrier into the surface, captured by the electroneutral surface states. On one hand, the captured electrons can form a negative electric field to deplete the channel electrons, and then a negative electric potential ($V_e$) is provided by the tunneled electrons; On the other hand, the tunneled electrons can decrease the 2DEG electrons in the channel. The number of decreased 2DEG electrons in the subthreshold region is significant because it can effectively affect the channel current, namely the $I_{ds}$.

**Fig. 3(a)** showed the semi-log transfer curves with forward/reverse sweep in $V_{gs}$. Here we choose the three parts (*a~b*, *b~c*, *d~e*) to explain the SS variation in the subthreshold regions. At *a~b* region, $V_{gs}$ is smaller than the threshold voltage $V_{th}$. As shown in **Fig. 3(b)**, the gate channel region is depleted by the gate voltage. There are few 2DEG electrons under the gate region. Therefore, few channel electrons can tunnel to the surface. With the increased forward $V_{gs}$ (at *b~c* region), the device begins to turn on and the 2DEG electrons starts to inject onto the surface, as shown in **Fig. 3(c)**. Here SS between *b* and *c* can be written as SS = $\partial V_{gs}/\partial \log_{10}(I_{ds})$ = $((V_{gsc}+V_e) - V_{gsb})/(\log_{10}(I_{dsc})-\log_{10}(I_{dsb})) = (V_{gsc} - V_{gsb} + V_e)/(\log_{10}(I_{dsc}) - \log_{10}(I_{dsb}))$, where $V_{gsc}$, $V_{gsb}$, $I_{dsc}$, and $I_{dsb}$ are the gate-source voltage and drain-source current at point *b* and *c*, respectively. Here $V_{gsc} > V_{gsb}$ and $V_e$ is negative, therefore $(V_{gsc} - V_{gsb} + V_e)$ decreased compared to that without electron injections (namely, $V_{gsb} - V_{gsc}$). The electrons tunnel from 2DEG channel to the surface, resulting in the decrease of channel electron and channel current. So $I_{dsc}$ decreased compared to that without electron injections, leading to the decreasing of ($\log_{10}(I_{dsc})$-$\log_{10}(I_{dsb})$). Both ($V_{gsc} - V_{gsb} + V_e$) and ($\log_{10}(I_{dsc})$-$\log_{10}(I_{dsb})$) decrease and therefore it's hard to judge the decrease or increase of SS. Conversely, with the decreased $V_{gs}$ (reverse, from *d* to *e*), the electrons start to release from surface to the channel. SS can be written as SS = $\partial V_{gs}/\partial \log_{10}(I_{ds}) = (V_{gse} - (V_{gsd} + V_e))/(\log_{10}(I_{dse})-\log_{10}(I_{dsd})) = (V_{gse} - V_{gsd} - V_e)/(\log_{10}(I_{dse})-\log_{10}(I_{dsd}))$, where $V_{gse}$, $V_{gsd}$, $I_{dse}$, and $I_{dsd}$ are the gate-source voltage and drain-source current at point *e* and *d*, respectively. $V_{gse} < V_{gsd}$ and $V_e$ is negative, therefore ($V_{gse} - V_{gsd} - V_e$) is smaller than that without electrons release. The electron release increased the channel electron and channel current. $\log_{10}(I_{dsd})$ decreased compared to that without electron injection. The decreased ($V_{gse} - V_{gsd} - V_e$) and increased ($\log_{10}(I_{dse})$-$\log_{10}(I_{dsd})$), leading to a steep SS below 60 mV/dec.

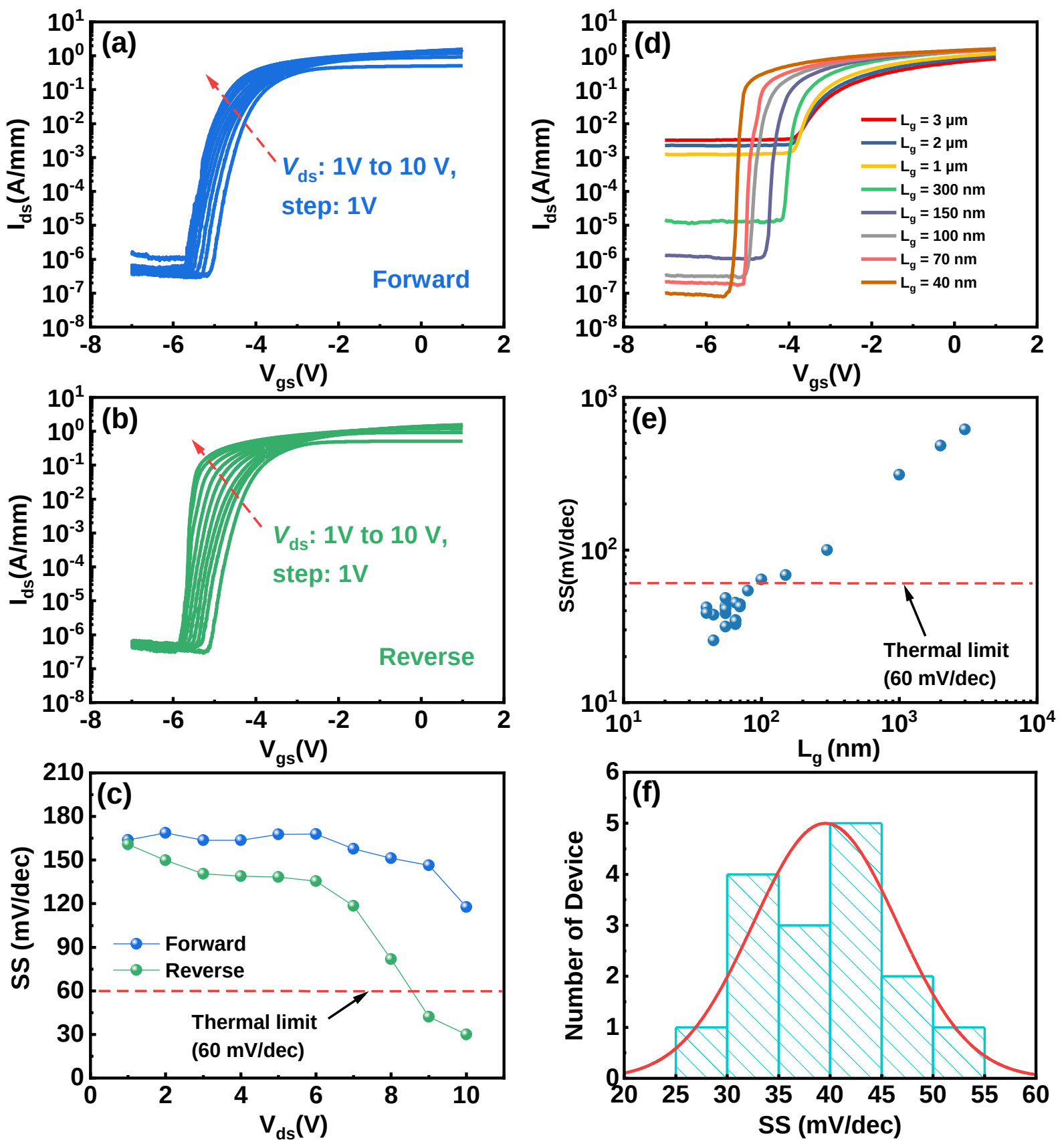

**Fig. 4** The transfer curves of 40-nm InAlN/GaN HEMT in semi-log at $V_{ds}$ changed from 1 V to 10 V with 1 V step **(a)** with forward sweep in $V_{gs}$ and **(b)** with reverse sweep in $V_{gs}$. **(c)** The average SS as a function of $V_{ds}$ with forward/reverse sweep in $V_{gs}$. **(d)** The transfer curves in semi-log with different gate length ($L_g$) at $V_{ds}$ = 10 V. (e) The average SS as a function of $L_g$. When $L_g$ < 100 nm, the SS falls below 60 mV/dec. (f) The numbers of fabricated devices with SS below 60 mV/dec.

To confirm this explanation, the transfer curves of the 40-nm InAlN/GaN HEMT at different $V_{ds}$ (from 1 V to 10 V, with 1 V step) were measured and shown in **Fig. 4(a)** and **(b)**. It is observed that the $I_{ds}$-$V_{ds}$ slope became steeper with increased $V_{gs}$ in both sweep directions. **Fig. 4(c)** shows the extracted average SS as a function of $V_{ds}$. It is shown that SS decreased with increased $V_{ds}$ with forward/reverse sweep in $V_{gs}$. In forward sweep, SS decreased from 164 mV/dec (at $V_{ds}$ = 1V) to 118 mV/dec (at $V_{ds}$ = 10 V). In reverse sweep, SS decreased from 161 mV/dec at $V_{ds}$ = 1 V to 30 mV/dec(at $V_{ds}$ = 10 V). With the increased $V_{ds}$, the SS difference between in forward direction and in reverse direction became more pronounced. The increased $V_{ds}$ can increase the 2DEG electron energy and increase the electron injection to the surface, which can decrease SS. Therefore, with increased $V_{ds}$, SS is decreased.

**Fig. 4(d)** shows the measured transfer curves of devices with different $L_{gs}$. When $L_g$ decreased from 3 μm to 40 nm, the off current ($I_{off}$) decreased from $3.26 \times 10^{-3}$ A/mm ($L_g$ = 3 μm) to $9.96 \times 10^{-8}$ A/mm ($Lg$ = 40 nm), and on/off current ($I_{on}/I_{off}$) ratio increased from $2.51 \times 10^2$ ($L_g$ = 3 μm) to $1.59 \times 10^7$ ($Lg$ = 40 nm). This means the high energy electron injection can effectively deplete the channel electrons, decrease $I_{off}$, and increase $I_{on}/I_{off}$ ratio, which results in the improved device subthreshold swing. **Fig. 4(e)** shows the extracted average SS with different $L_g$. SS shows a significantly decrease with the decreased $L_g$ and falls below 60 mV/dec when $L_g$ < 100 nm. All these devices are fabricated on the same chip and with the same fabrication technology. With the gate length scales down, the electric field in the source-drain channel increases. The 2DEG electron velocity increases under the increased electric field and the electron energy increases, leading to the electrons injection and decreased SS. **Fig. 4(f)** shows the number of devices with SS below 60 mV/dec. The total number of devices is 16. All of gate lengths are below 100 nm, confirming the electron injection behavior existing in all down-scaled devices.

In conclusion, an average SS of 30 mV/dec over three orders of magnitude in $I_{ds}$ and a minimum point-by-point SS of 15 mV/dec were achieved in the 40-nm InAlN/GaN HEMTs. It is found that SS decreases with $V_{ds}$ as well as $L_g$, and falls below 60 mV/dec when $L_g$ < 100 nm. The decreasing of SS as the device dimension scales down is attributed to the tunneling of high energy electrons from channel to the surface. This shows that the great potential of the InAlN/GaN HEMTs to be applied in future logic switches.

References

[1] Asif Islam Khan, Korok Chatterjee, Brian Wang, Steven Drapcho, Long You, Claudy Serrao, Saidur Rahman Bakaul, Ramamoorthy Ramesh, and Sayeef Salahuddin, Nat. Mater. **14** (2), 182 (2015).

[2] Felicia A McGuire, Yuh-Chen Lin, Katherine Price, G Bruce Rayner, Sourabh Khandelwal, Sayeef Salahuddin, and Aaron D Franklin, Nano Lett. **17** (8), 4801 (2017).

[3] Sayeef Salahuddin and Supriyo Datta, Nano Lett. **8** (2), 405 (2008).

[4] Eng-Huat Toh, Grace Huiqi Wang, Ganesh Samudra, and Yee-Chia Yeo, J. Appl. Phys. **103** (10), 104504 (2008).

[5] Woo Young Choi, Byung-Gook Park, Jong Duk Lee, and Tsu-Jae King Liu, IEEE Electron Device Lett. **28** (8), 743 (2007).

[6] Asif I Khan, Chun W Yeung, Chenming Hu, and Sayeef Salahuddin, presented at the 2011 International Electron Devices Meeting, 2011, pp. 11.3. 1.

[7] Asif Islam Khan, Korok Chatterjee, Juan Pablo Duarte, Zhongyuan Lu, Angada Sachid, Sourabh Khandelwal, Ramamoorthy Ramesh, Chenming Hu, and Sayeef Salahuddin, IEEE Electron Device Lett. **37** (1), 111 (2015).

[8] Yi-Feng Wu, David Kapolnek, James P Ibbetson, Primit Parikh, Bernd P Keller, and Umesh K Mishra, IEEE Trans. Electron Devices **48** (3), 586 (2001).

[9] N Tipirneni, Alexei Koudymov, V Adivarahan, Jinwei Yang, Grigory Simin, and M Asif Khan, IEEE Electron Device Lett. **27** (9), 716 (2006).

[10] Daisuke Shibata, Ryo Kajitani, Masahiro Ogawa, Kenichiro Tanaka, Satoshi Tamura, Tsuguyasu Hatsuda, Masahiro Ishida, and Tetsuzo Ueda, presented at the 2016 IEEE International Electron Devices Meeting (IEDM), 2016, pp. 10.1. 1.

[11] HW Then, S Dasgupta, M Radosavljevic, L Chow, B Chu-Kung, G Dewey, S Gardner,

X Gao, J Kavalieros, and N Mukherjee, presented at the 2013 IEEE International Electron Devices Meeting, 2013, pp. 28.3. 1.

[12] Qi Zhou, Sen Huang, Hongwei Chen, Chunhua Zhou, Zhihong Feng, Shujun Cai, and Kevin J Chen, presented at the 2011 International Electron Devices Meeting, 2011, pp. 33.4. 1.

[13] Bo Song, Mingda Zhu, Zongyang Hu, Meng Qi, X Yan, Yu Cao, Erhard Kohn, Debdeep Jena, and Huili Grace Xing, presented at the 2014 Silicon Nanoelectronics Workshop (SNW), 2014, pp. 1.

[14] Zongyang Hu, Raj Jana, Meng Qi, Satyaki Ganguly, Bo Song, Erhard Kohn, Debdeep Jena, and Huili Grace Xing, presented at the 72nd Device Research Conference, 2014, pp. 27.

[15] Yuanzheng Yue, Zongyang Hu, Jia Guo, Berardi Sensale-Rodriguez, Guowang Li, Ronghua Wang, Faiza Faria, Tian Fang, Bo Song, and Xiang Gao, IEEE Electron Device Lett. **33** (7), 988 (2012).

[16] Peng Cui, Andrew Mercante, Guangyang Lin, Jie Zhang, Peng Yao, Dennis W Prather, and Yuping Zeng, Appl. Phys. Express **12** (10), 104001 (2019).